\documentclass[conference]{IEEEtran}

\usepackage[utf8]{inputenc}
\usepackage[english]{babel}
\usepackage{graphicx}
\usepackage{psfrag}
\usepackage{amsmath}
\usepackage[english,algoruled,commentsnumbered,onelanguage,vlined]{algorithm2e}

\begin{document}

\title{Decoding Superposed LoRa Signals}

\author{
\IEEEauthorblockN{Nancy El Rachkidy$^{(1)}$, Alexandre Guitton$^{(1)}$, Megumi Kaneko$^{(2)}$}
\IEEEauthorblockA{
(1) Université Clermont Auvergne, CNRS, LIMOS, F-63000 Clermont-Ferrand, France\\
(2) National Institute of Informatics, Hitotsubashi, 2-1-2, Chiyoda-ku, 101-8430 Tokyo, Japan\\
Emails: nancy.el\_rachkidy@uca.fr, alexandre.guitton@uca.fr, megkaneko@nii.ac.jp
}
}

\maketitle

\begin{abstract}
Long-range low-power wireless communications, such as LoRa, are used in many IoT and environmental monitoring applications. They typically increase the communication range to several kilometers, at the cost of reducing the bitrate to a few bits per seconds. Collisions further reduce the performance of these communications. In this paper, we propose two algorithms to decode colliding signals: one algorithm requires the transmitters to be slightly desynchronized, and the other requires the transmitters to be synchronized. To do so, we use the timing information to match the correct symbols to the correct transmitters. We show that our algorithms are able to significantly improve the overall throughput of LoRa.
\end{abstract}


\IEEEpeerreviewmaketitle

\section{Introduction}
\label{section:introduction}

Long-range low-power communication technologies such as LoRa~\cite{lora}, Sigfox~\cite{sigfox}, and Ingenu~\cite{ingenu}, are becoming widely used in Low-Power Wide Area Networks (LPWANs). These technologies are suitable to cover large zones and are thus becoming attractive technologies for Internet of Things (IoT) and monitoring applications~\cite{centenaro16long,nolan16evaluation,petajajarvi17evaluation}.

LoRa~\cite{lora} is a recent physical layer for LPWANs, which extends the communication range by reducing the throughput. LoRaWAN~\cite{lorawan-2017} is a simple MAC protocol based on LoRa, which allows end-devices (ED) to communicate with a small duty-cycle (1\%) to a network server through gateways. Thus, EDs can save energy, and the network lifetime is increased.

The main issue in LoRa and LoRaWAN is the limited throughput: the indicative physical bitrate varies between 250 and 11000 bps~\cite{lorawan-regional-settings-2017}. Moreover, when two EDs transmit simultaneously using the same parameters, a collision occurs and none of the signals are decoded by LoRa. Thus, both EDs have to retransmit, which further reduces the throughput.

In this paper, we show that it is possible to retrieve the frames from superposed signals. For the case where superposed signals are slightly desynchronized, we propose a linear algorithm based on timing information that attempts to decode all frames. This algorithm always succeeds when there are two signals. We prove that for three or more signals, it is not always possible to decode each signal. Next, for the case where superposed signals are completely synchronized, we propose a simple algorithm requiring only one retransmission to deduce the other colliding frame. To the best of our knowledge, this is the first work on LoRa interference cancellation.

The remainder of this paper is as follows. Section~\ref{section:state-of-the-art} describes the modulation of LoRa. Section~\ref{section:proposition} describes our two cases (slightly desynchronized and completely synchronized), and presents our two algorithms. Section~\ref{section:results} gives our simulation results. Finally, Section~\ref{section:conclusion} concludes the paper.

\section{State of the art}
\label{section:state-of-the-art}

In the following, we first describe the MAC protocol LoRaWAN, and then the physical layer LoRa. Note that our paper proposes an improvement to LoRa, which can be used to improve the performance of any MAC protocol based on LoRa (including LoRaWAN).

\subsection{LoRaWAN}

LoRaWAN (in version 1.0~\cite{lorawan-2015} or in version 1.1~\cite{lorawan-2017}) is a MAC protocol based on LoRa. Three classes are defined, depending on the communication paradigm: Class A is for low-power uplink communications, Class B is for delay-guaranteed downlink communications, and Class C is for EDs without energy constraints. In Class A, the only mandatory class of LoRaWAN, an ED can transmit at any time. It chooses a channel randomly, sends the frame, and waits for an acknowledgment during two successive receive windows. After its transmission, the ED is forbidden to transmit for a delay equal to 99 times the duration of the frame transmission. In this way, the transmission time of EDs does not exceed 1\%.

LoRaWAN adapts the bitrate to the quality of links by implementing a trade-off between the signal robustness and the bitrate, through the use of the Spreading Factor (SF) of the signal: when an ED experiences a low signal quality, it increases its SF, which results into lower bitrate, but better decoding capabilities of the signal. This modification is controlled by the datarate (DR) of LoRaWAN, which is a value ranging from DR0 (large SF, small bitrate) to DR6 (small SF, large bitrate).

European regional settings of LoRaWAN~\cite{lorawan-regional-settings-2017} define most LoRa parameters. The bandwidth of channels, $BW$, is equal to 125 kHz for DR0 to DR5, and 250 kHz for DR6. The SF varies from 12 down to 7 for DR0 to DR5, and is equal to 7 for DR6. The preamble length is equal to 8 symbols. The physical bitrate varies between 250 bps for DR0, to 11000 bps for DR6. The maximum MAC payload of a frame varies between 59 bytes for DR0 and 230 bytes for DR6.

\subsection{LoRa}

LoRa is a physical layer technology for LPWAN, based on a Chirp-Spread Spectrum (CSS) modulation. In this modulation, each LoRa chirp consists of a linear frequency sweep. The duration of the sweep is called symbol duration (SD), and depends on the value of SF. The sweep is performed over a frequency range of size $BW$. Chirps are divided into up-chirps, where the frequency sweep is increasing, and down-chirps, where the frequency sweep is decreasing.

\begin{figure}[htbp]
    \centering
    \psfrag{frequency}[c]{{\small frequency}}
    \psfrag{BW}[c]{{\small $BW$}}
    \psfrag{value}[c]{{\small value}}
    \psfrag{SD}[c]{{\small $SD$}}
    \psfrag{time}[c]{{\small time}}
    \includegraphics[scale=0.5]{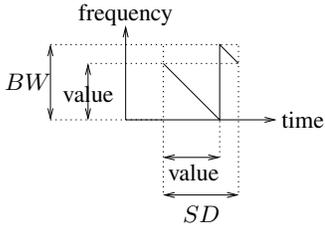}
    \caption{Example of a single LoRa down-chirp. Computing the symbol value requires identifying the sharp frequency edge.}
    \label{figure:lora-one-symbol}
\end{figure}

Each chirp can encode $2^{SF}$ symbol values. To do this, LoRa shifts the sweep by the symbol value, as shown on Fig.~\ref{figure:lora-one-symbol} for a down-chirp. The receiver is able to detect the sharp edge in the instantaneous frequency trajectory~\cite{goursaud15dedicated}. The symbol value is equal to the shift in the frequency at the beginning of the symbol. It is also proportional to the time between the beginning of the symbol and the sharp frequency edge. For up-chirps, it is proportional to the remaining time between the sharp frequency edge and the end of the symbol.

To decode the value of a symbol, the receiver needs to know the frontier of the symbol. Thus, LoRa introduces a preamble of a few symbols (typically, eight). In uplink communications, the preamble consists of up-chirps and the data consists of down-chirps. In downlink communications, the preamble consists of down-chirps and the data consists of up-chirps.

\begin{figure}[htbp]
    \psfrag{sender}[c]{{\small sender}}
    \psfrag{receiver}[c]{{\small receiver}}
    \psfrag{preamble}[c]{{\small preamble}}
    \psfrag{s1}[c]{{\small 3}}
    \psfrag{s2}[c]{{\small 0}}
    \psfrag{s3}[c]{{\small 2}}
    \psfrag{s4}[c]{{\small 2}}
    \psfrag{desynchronization information}[c]{{\small desynchronization information}}
	\includegraphics[scale=0.5]{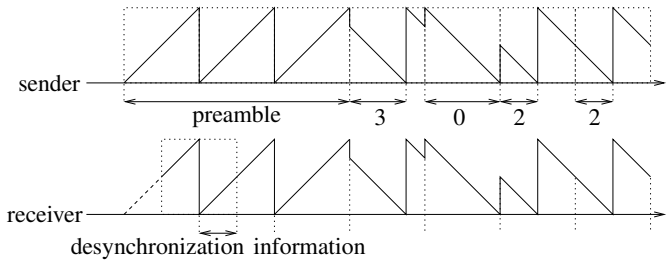}
	\caption{Example of a LoRa uplink frame, with a short preamble and four symbols of data, with $SF=2$. The receiver synchronizes itself with the sender during the preamble.}
	\label{figure:lora-one-signal}
\end{figure}

Figure~\ref{figure:lora-one-signal} shows an example of an uplink communication with a short preamble (three symbols) and a few data symbols (four symbols). We chose $SF=2$ for the sake of simplicity, leading to $2^{SF}=4$ possible values per symbol. Let us assume that a desynchronized node starts receiving the preamble, not necessarily at the exact beginning of the preamble. The node detects a sharp frequency edge, which indicates the frontier of a symbol. From this information, the receiver can synchronize itself according to the transmitter. The end of the preamble is detected by the inversion of the chirps. In this example, the data symbols are 3, 0, 2, 2.

\section{Cancellation of LoRa signals}
\label{section:proposition}

LoRa gateways are able to decode superposed LoRa signals as long as they are sent on different SFs. Notice however that some researchers have shown that signals on different SFs are not completely orthogonal~\cite{croce17impact,zhu18evaluation}.

When several signals are received on the same channel and with the same SF, a difference of received power might cause the strongest signal to be captured~\cite{goursaud15dedicated,haxhibeqiri17lora}. When several signals have a similar receive power, a collision occurs and all signals are considered lost.

In this paper, we focus on decoding superposed LoRa signals of {\it similar receive power}, on the {\it same channel}, with the {\it same SF}. To do so, we show that we can use timing information to match the correct symbols to the correct ED.

In Subsection~\ref{subsection:assumptions}, we describe our assumptions. In Subsection~\ref{subsection:two-signals}, we provide our main algorithm, and we describe how it can decode two signals that are slightly desynchronized. In Subsection~\ref{subsection:three-signals}, we extend the discussion for the case of three signals (or more) that are slightly desynchronized. Finally, in Subsection~\ref{subsection:discussion}, we present a simple algorithm for the case of signals that are completely synchronized.

Note that our algorithms cannot be applied directly on LoRaWAN, as most communications in LoRaWAN are not synchronized. However, our algorithms could enable the design of a novel synchronized MAC layer based on LoRa, tailored to the star like topology of LoRaWAN, to reach better performances than the basic LoRaWAN.

\subsection{Assumptions}
\label{subsection:assumptions}

We assume that there are no non-linearity effects between down-chirps (respectively up-chirps). In other words, if two down-chirps (resp. up-chirps) $c_1$ and $c_2$ overlap at a given time $t$ at the receiver side, the two observed frequencies are the frequency of $c_1$ (at time $t$) and the frequency of $c_2$ (at time $t$). Without additional information, it is not possible to identify the correct frequency of each transmitter. We assume that when an up-chirp is superposed with a down-chirp, it is not possible to detect any of the frequencies.

We also assume that it is possible for the hardware of the receiver to detect all frequencies of overlapping down-chirps (resp. up-chirps) within $\delta$ time-units. In the following examples, we will use $\delta=SD/4$ unless stated otherwise\footnote{Please note that on real LoRa hardware, the decoding of signals is not carried out by directly detecting the sharp frequency edges, but instead by computing a fast Fourier transform and detecting the peak in the frequency domain~\cite{goursaud15dedicated}. With our proposition, this translates into either detecting the two sharp frequency edges in the time domain, or the two peaks in the frequency domain.}.

We assume that when several frequencies overlap at a given time, only one frequency is detected by the receiver. For instance, if there are three nodes transmitting at a given time, but only two frequencies $f_1$ and $f_2$ are detected, we assume that it is not possible to know whether two nodes were transmitting with $f_1$ and one with $f_2$, or one node was transmitting with $f_1$ and two with $f_2$.

We also assume some properties on the frames: all nodes transmit with the same preamble duration, the frame length is included at the beginning of the frame, and there is at least one symbol change during the whole frame (that is, a data payload does not consist of a sequence of identical symbols).

Finally, in the following, we consider two cases: the case where nodes are slightly desynchronized, and the case where nodes are fully synchronized. In the case where nodes are slightly desynchronized, we assume that all nodes start their transmission within $SD-\delta$ time units, and that any two nodes start their transmission with a delay of $\delta$ time units or more. In the following examples, if there are three transmitting nodes, we consider that node $n_1$ starts at time $t_0$, node $n_2$ at time $t_0+\delta$, and node $n_3$ at time $t_0+2\delta$.

\subsection{Case of two slightly desynchronized signals}
\label{subsection:two-signals}

In this subsection, we consider the superposition of two signals from two transmitters that are slightly desynchronized (by at least $\delta$ time units, and at most $SD-\delta$ time units).

\begin{figure}[htbp]
	\centering
    \psfrag{n1}[c]{{\small $n_1$}}
    \psfrag{n2}[c]{{\small $n_2$}}
    \psfrag{receiver}[c]{{\small receiver}}
    \psfrag{t0}[c]{{\small $t_0$}}
    \psfrag{t1}[c]{{\small $t_1$}}
    \psfrag{t2}[c]{{\small $t_2$}}
    \psfrag{t3}[c]{{\small $t_3$}}
    \psfrag{t4}[c]{{\small $t_4$}}
    \psfrag{t5}[c]{{\small $t_5$}}
    \psfrag{t6}[c]{{\small $t_6$}}
    \psfrag{t7}[c]{{\small $t_7$}}
    \psfrag{t8}[c]{{\small $t_8$}}
    \psfrag{t9}[c]{{\small $t_9$}}
    \psfrag{t10}[c]{{\small $t_{10}$}}
    \psfrag{t11}[c]{{\small $t_{11}$}}
    \psfrag{t12}[c]{{\small $t_{12}$}}
    \psfrag{0}[c]{{\small 0}}
    \psfrag{1}[c]{{\small 1}}
    \psfrag{2}[c]{{\small 2}}
    \psfrag{3}[c]{{\small 3}}
    \psfrag{delta}[c]{\small $\delta$}
	\includegraphics[scale=0.5]{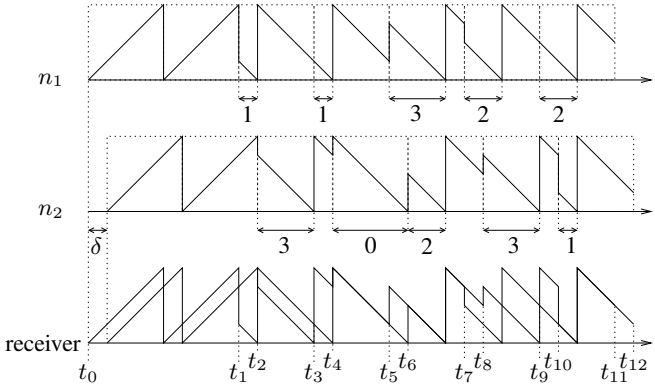}
	\caption{Superposition of two slightly desynchronized signals.}
	\label{figure:lora-two-signals}
\end{figure}

Figure~\ref{figure:lora-two-signals} shows an example of the reception of two slightly desynchronized signals. The preamble length is two symbols, and $SF=2$. The figure shows the signal of the first transmitter $n_1$ starting at $t_0$, the signal of the second transmitter $n_2$ starting at $t_0+\delta$, and the superposed signal at the receiver. Note that the data transmitted by $n_1$ is $(1,1,3,2,2)$, and the data transmitted by $n_2$ is $(3,0,2,3,1)$. We will first explain our algorithm on this example, and then proceed with a more formal description.

\underline{\it Example of Preamble detection and data decoding}

In this paragraph, we give an example in order to explain how the receiver, using our proposition, can detect preambles and decode data sent by two transmitters.

{\it Preamble detection:} During $[t_0;t_0+\delta]$, the receiver detects the preamble of $n_1$. During $[t_0+\delta;t_0+2\delta]$, the receiver is able to detect that two slightly desynchronized signals are transmitted, and is able to deduce the symbol frontiers of both transmitters. At frontier $t_1$, or more precisely, during $[t_1;t_1+\delta]$, the receiver is not able to detect the superposition of preambles anymore (mixed up and down chirps). Thus, it knows that the transmission of the first data symbol of $n_1$ has started. This symbol is currently undecodable due to the overlapping of an up-chirp with a down-chirp.

{\it Data decoding:} We define the sequence of decoded data for $n_1$ by $d_1$ and the sequence of decoded data for $n_2$ by $d_2$. At frontier $t_2$, the receiver stores the current frequencies, which correspond to $F_{lim_-}(t_2)=\{0,3\}$. At frontier $t_3$, the receiver computes $F_{lim_+}(t_3)$ by updating the previous frequencies $F_{lim_-}(t_2)=\{0,3\}$, and obtains $F_{lim_+}(t_3)=\{0,1\}$ (each frequency is reduced by 3 since $3\delta$ time units have passed since $t_2$, as $\delta=SD/4$). It detects the current frequencies $F_{lim_-}(t_3)=\{0,1\}$. There is no change in the frequencies ($F_{lim_+}(t_3)=F_{lim_-}(t_3)$), since the beginning of the data of $n_1$ starts with the repeated symbol 1. Thus, the algorithm leaves $*$ for the first symbol of $n_1$ (to be decoded later), so $d_1=(*,1)$. At frontier $t_4$, the receiver computes $F_{lim_+}(t_4)$ by updating the previous frequencies $F_{lim_-}(t_3)=\{0,1\}$, and obtains $F_{lim_+}(t_4)=\{0,3\}$ (since $\delta$ time units have passed). It detects the current frequencies $F_{lim_-}(t_4)$, and obtains $F_{lim_-}(t_4)=\{0\}$, which is equivalent to $\{0,0\}$. Thus, one frequency changed from 3 to 0, which is that of $n_2$, since it is a frontier of $n_2$, hence, $d_2=(3,0)$. Thus, the current symbol of $n_1$ corresponds to frequency 0 too (which is translated into 1 at the beginning of the symbol frontier of $n_1$, which was $t_3$). At frontier $t_5$, the receiver computes $F_{lim_+}(t_5)$ by updating the previous frequencies $F_{lim_-}(t_4)=\{0,0\}$, and obtains $F_{lim_+}(t_5)=\{1,1\}$. It detects the current frequencies $F_{lim_-}(t_5)=\{1,3\}$. The frequency of $n_1$ changed from 1 to 3, hence $d_1=(*,1,3)$. So, the current symbol of $n_2$ corresponds to frequency 1 (which is translated to 0 at the beginning of the symbol frontier of $n_2$, which is $t_4$). The algorithm continues until $t_{12}$, where no frequency is received. Thus, the algorithm knows that all nodes have stopped their transmissions. The algorithm removes the last predicted symbol of $n_1$ (indeed, at $t_{11}$, it considered that $n_1$ was transmitting a symbol with the same frequency as the frequency of $n_2$). At this step, the decoded frames are $d_1=(*,1,3,2,2)$ for $n_1$ and $d_2=(3,0,2,3,1)$ for $n_2$. Then, the algorithm replaces all special values $*$ with the first known value of the frame. The algorithm uses the frame length present in each frame to truncate the frames to their correct length. Finally, the algorithm outputs are $(1,1,3,2,2)$ and $(3,0,2,3,1)$, as expected.

\underline{\it Generalization of Preamble detection and data decoding}

In this paragraph, we generalize the example given above and we formulate our proposition in Algorithm~\ref{algorithm:proposition}. 

{\it Preamble detection:} The superposition of preambles will result in the superposition of up-chirp symbols, except for the end of the last preamble. Thus, the receiver will detect two sharp frequency edges for most preamble symbols. Each of this sharp edge will allow the receiver to know the symbol frontier of a transmitter. The beginning of the first data symbol of the first node is not decodable, as it corresponds to a down-chirp superposed with the up-chirp of the end of the preamble of the second node.

{\it Data decoding:} From the first data symbol of the second node, only down-chirps are superposed, and thus it is possible to detect all sharp edges. The difficulty relies in correlating each frequency with the symbols of each node. To do so, we use the following property: sharp edges can occur only at the beginning of a symbol, when the symbol changes, or once during a symbol. When the sharp edge occurs during a symbol, it can be predicted if the symbol value is known.

Algorithm~\ref{algorithm:proposition} describes our proposed algorithm. It starts after the superposed preambles have been received, and thus considers that the symbol frontier of each transmitter is known. The algorithm considers the frontiers of all data symbols sequentially, apart from the first frontier of the first node which cannot be decoded. At each frontier, the receiver updates the previous frequencies (since frequencies change over time in LoRa chirps, and time has passed since the detection of the previous frequencies). Then, the receiver compares these (updated) previous frequencies with the current frequencies\footnote{In practice, it may take up to $\delta$ time units to obtain the current frequencies, so the receiver might have to update the current frequencies based on the detection time.}. The following two cases are the only possible cases.\\
{\it Case 1:} One frequency has changed. This can only happen when a new symbol starts, which can only occur at the symbol frontier. Since the receiver knows if the current frontier is for the first or the second transmitter, it knows the new symbol for the current node (based on the new frequency), the previous symbol for the current node (based on the frequency that has changed), and the current symbol for the other node (based on the frequency that did not change).\\
{\it Case 2:} No frequency has changed. This can only happen when the new symbol is equal to the previous symbol (this was the case on Fig.~\ref{figure:lora-two-signals} at times $t_3$ and $t_9$). If the receiver knows the previous symbol of the current node (time $t_9$ of Fig.~\ref{figure:lora-two-signals}), the new symbol can be deduced. Note that at the beginning of the algorithm, however, the first symbol value cannot be deduced when it is repeated (time $t_3$ of Fig.~\ref{figure:lora-two-signals}). In this case, the algorithm leaves a special value (denoted by $*$ here). As soon as one symbol changes, the receiver is able to deduce the values of all these repeated symbols. This is why we assumed at least one symbol change per frame.

\begin{algorithm}[htbp]
    \For{each frontier $t_i$ of a data chirp}{
        compute currentSymbol and currentNode\\
        \If{currentSymbol=0 and currentNode=1}{
            skip (frequencies cannot be detected)\\
        }
        \Else{
            $F_{lim_-}(t_i)\leftarrow$detect current frequencies\\
            \If{currentSymbol=0 and currentNode=2}{
                skip ($F_{lim_-}(t_i)$ is already computed)\\
            }
            \Else{
                compute $F_{lim_+}(t_{i})$ by updating $F_{lim_-}(t_{i-1})$\\
                $changedF\leftarrow{}F_{lim_-}(t_i)-F_{lim_+}(t_i)$\\
                \If{$|changedF|=0$}{
                    the new symbol in $symb[$currentNode$]$ is equal to the previous (or $*$)
                }
                \Else{
                    the previous symb. in $symb[$currentNode$]$ is equal to the value of $F_{lim_+}(t_i)-F_{lim_-}(t_i)$\\
                    the new symbol in $symb[$currentNode$]$ is equal to the value of $changedF$
                }
            }
        }
    }
    \For{each node $n$}{
        replace in $symb[n]$ all the leading $*$ values to the first defined value\\
        truncate the frame according to its length
    }
    \caption{Decoding of two slightly desynchronized superposed LoRa signals.}
    \label{algorithm:proposition}
\end{algorithm}

The time complexity of our algorithm is linear with the number of symbols of the longest frame. Most of the symbols are decoded on the fly, $\delta$ time units after the beginning of the symbol, except for the symbols repeated initially (see the last loop of the algorithm). The space complexity of our algorithm is $\mathcal{O}(1)$, since the storage requirement is limited to the value of the first non-special symbol for each node. Thus, the algorithm is optimal in time and space, for two nodes\footnote{As we will see in Subsection~\ref{subsection:three-signals}, our algorithm is not able to decode all frames for three nodes or more, so it cannot be considered optimal in this case.}.

\subsection{Case of three slightly desynchronized signals}
\label{subsection:three-signals}

\begin{figure}[htbp]
	\centering
    \psfrag{n1}[c]{{\small $n_1$}}
    \psfrag{n2}[c]{{\small $n_2$}}
    \psfrag{n3}[c]{{\small $n_3$}}
    \psfrag{receiver}[c]{{\small receiv.}}
    \psfrag{0}[c]{{\small 0}}
    \psfrag{1}[c]{{\small 1}}
    \psfrag{2}[c]{{\small 2}}
    \psfrag{3}[c]{{\small 3}}
	\includegraphics[scale=0.5]{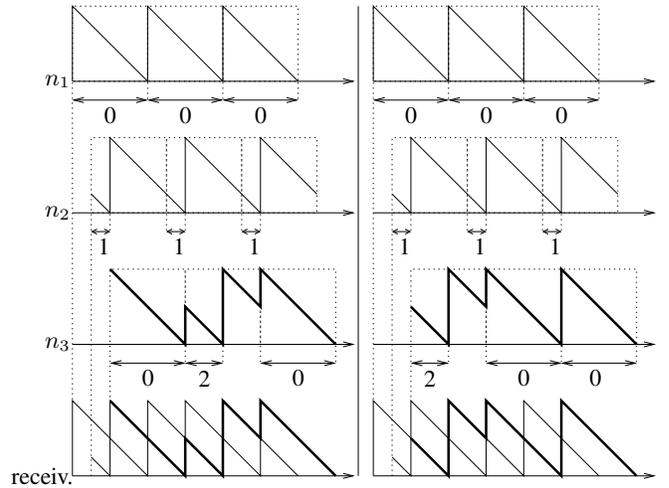}
	\caption{When three nodes that are slightly desynchronized transmit frames, it is not always possible to decode them: these two sets of frames produce the same superposition of frequencies.}
	\label{figure:undecodable}
\end{figure}

Note that with our hypotheses, decoding three or more signals is not always possible. For instance, Fig.~\ref{figure:undecodable} shows two sets of different signals that produce the same superposition of frequencies, and thus cannot be decoded.

Our algorithm is able to decode many cases of slightly desynchronized signals for $n$ transmitters, when $n\geq{}3$. It only fails to do so when the number of received frequencies is within $[2;n-1]$ (which never occurs when $n=2$). In this case, even if the algorithm knows that the frequency of the current node has changed, it cannot determine what is the new value, as it has $n-1>1$ possibilities. It can still deduce the value of the previous symbol for this node. At the next frontier for this node, though, the value of this symbol might be deduced, depending on the number of other frequencies.

\subsection{Case of two synchronized signals}
\label{subsection:discussion}

We now consider the case of two synchronized signals.

When the two transmitters are completely synchronized, it can be noticed that (at most) two values for each symbol duration are obtained, one for each transmitter. With our assumptions, though, it is not possible to match each value to the correct transmitter. The uncertainty of two values for each symbol might seem large, but it is quite small compared to the fact that each symbol carries in fact $SF$ bits of data.

Thus, we propose a simple algorithm for this case. When two such frames collide, the algorithm stores the possible values for each symbol, and requests any of the transmitters to retransmit its frame. When one frame is retransmitted, the algorithm is able to decode it, and is able to deduce the values of the colliding frame of the other node too, by elimination. Thus, instead of having to retransmit two colliding frames, only one retransmission is required.

\section{Numerical results}
\label{section:results}

In this section, we evaluate, by simulation, the performance of our algorithms in terms of percentage of successful decoding of colliding signals and throughput. We consider the two cases independently.

\subsection{Parameter Settings}
\label{subsection:settings}

Simulations are carried out using our own simulator developed in Perl. We model a network with a single gateway, a single network server, and one hundred EDs. We assume that all the EDs transmit on the same channel with the same SF, and that their signals are received at roughly the same power levels at the gateway, i.e., no capture conditions. We assume that time is divided into slots, and each ED has a probability $p$ to transmit a frame during a slot, with $p\leq{}0.01$ in order to be consistent with the duty-cycle of 1\%. For our algorithms, transmissions on the same slot are considered to be slightly desynchronized (in Subsection~\ref{subsection:desync}) or completely synchronized (in Subsection~\ref{subsection:sync}). We choose two values for SF: $SF7$ (which is the smallest SF in LoRaWAN) and $SF12$ (which is the largest SF in LoRaWAN). The frame length is set to $50$ bytes. We did not force frames to have at least one symbol change. However, the probability that a frame is generated with the same repeated symbol is very small, and we did not observe it during our simulations. Simulation results are obtained by averaging over ten thousand samples.

\subsection{Case of slightly desynchronized signals}
\label{subsection:desync}

\begin{figure}[htbp]
	\centering
	\psfrag{y}[c]{\%successful decoding}
	\psfrag{x}[c]{Number of superposed signals}
	\includegraphics[scale=0.875]{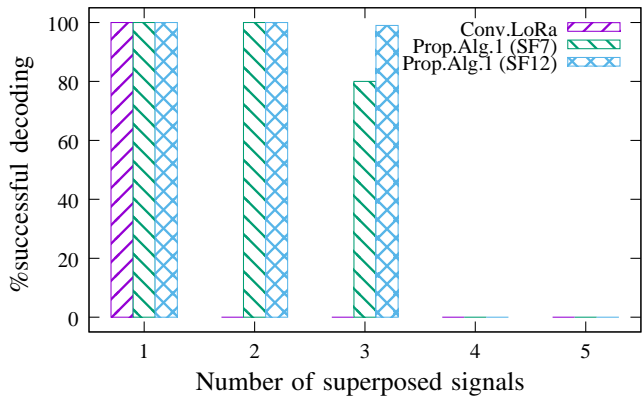}
	\caption{When the signals are slightly desynchronized, our algorithm is able to successfully decode some cases of colliding signals.}
	\label{figure:correction}
\end{figure}

Figure~\ref{figure:correction} shows the percentage of successful decoding of colliding signals, as a function of the number of colliding signals, in the case where signals are slightly desynchronized. When there are two or more colliding signals, LoRa is not able to decode any signal. When there are exactly two colliding signals, our algorithm is always able to decode both of them. When there are three or more colliding signals, our algorithm is not able to decode some signals: the proportion of signals that can be decoded depends on $SF$ and on the frame length. Indeed, when $SF$ is large, the number of possible values for each symbol is large, and the probability that several transmitters use the same frequency is low. For $SF=7$ and $n=3$ colliding frames, our algorithm is able to decode about 80\% of the frames. This number drops rapidly as the number of transmitters increases.

\begin{figure}[htbp]
	\centering
	\psfrag{y}[c]{\% Successful decoding}
	\psfrag{x}[c]{SF}
	\includegraphics[scale=0.875]{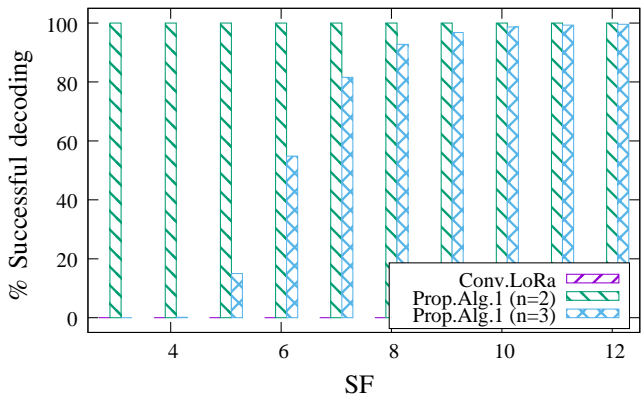}
	\caption{Our algorithm outperforms LoRa when a gateway receives $n=2$ or $n=3$ colliding signals, with the same SF and on the same channel.}
	\label{figure:correctionVsSF}
\end{figure}

Figure~\ref{figure:correctionVsSF} shows the percentage of successful decoding of colliding signals, as a function of the SF, for $n=2$ and $n=3$ superposed signals. We notice that LoRa is not able to decode colliding signals for any SF. This is due to the fact that in LoRa, a gateway cannot receive more than one signal on the same channel and with the same SF. However, we can see that our algorithm can decode both signals when $n=2$. When $n=3$, the performance of our algorithm varies significantly with SF. This is due to the fact that with a large SF, the probability to detect a single frequency decreases. Thus, the gateway increases its chances to receive a number of frequencies equal to the number of symbols to decode. Thus, compared to LoRa, our algorithm achieves a gain of 100\% when the gateway received two colliding signals and a gain between 18\% and 99\% when the gateway received three colliding signals.

\begin{figure}[htbp]
	\centering
	\psfrag{y}[c]{Throughput (bits/s)}
    \psfrag{x}[c]{Duty-cycle (in percentage)}
	\includegraphics[scale=0.875]{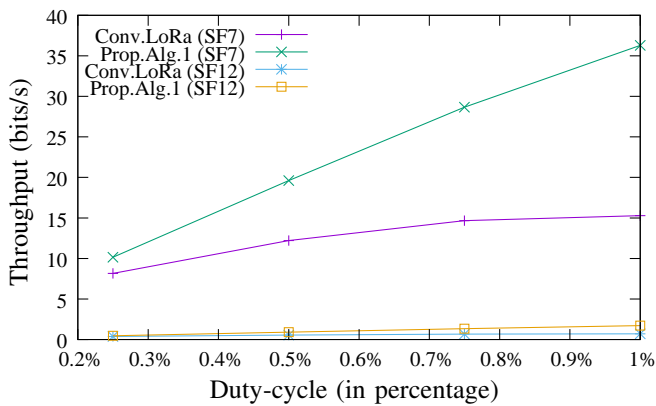}
	\caption{Colliding signals negatively impact the throughput in LoRa. However, our algorithm is able to increase the throughput up to 60\% when the duty-cycle is equal to 1\% and the network is of 100 EDs.}
	\label{figure:throughput}
\end{figure}

Figure~\ref{figure:throughput} shows the throughput as a function of the duty-cycle. In this scenario, we consider a network of one hundred EDs with a duty-cycle less or equal to 1\%. We compute throughput for LoRa and for our algorithm for two values of SF: SF7 and SF12. We notice that the throughput increases when the duty-cycle increases, since each ED sends more frames. Our algorithm enables to achieve a much higher throughput than LoRa, with a gain of up to 60\% for a duty-cycle of 1\%. This shows that our algorithm provides remarkable throughput gains, even at the system level.

\subsection{Case of synchronized signals}
\label{subsection:sync}

\begin{figure}[htbp]
	\centering
	\psfrag{y}[c]{\% Successful decoding}
	\psfrag{x}[c]{Number of colliding signals}
	\includegraphics[scale=0.875]{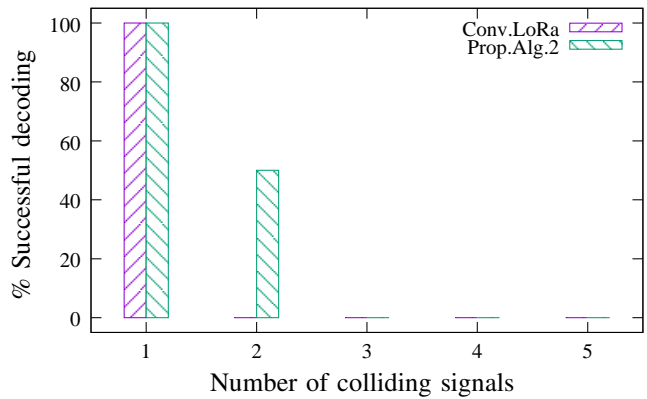}
    \caption{When the signals are synchronized, our algorithm is able to decode one frame per collision of two frames, provided that the other one is retransmitted.}
	\label{figure:correctionSync}
\end{figure}

Figure~\ref{figure:correctionSync} shows the percentage of successful decoding of colliding signals, as a function of the number of colliding signals, in the case where signals are synchronized. Since LoRa is unable to decode any signal when there are several EDs transmitting on the same channel and with the same SF, the percentage of successfully decoded signals is zero for two and more colliding signals. With our algorithm, when there are exactly two simultaneous transmissions, one of them can be decoded provided that one node retransmits its whole frame. Thus, for $n=2$ colliding signals, our algorithm decodes 50\% of the frames, accounting for the retransmission time.

\begin{figure}[htbp]
	\centering
	\psfrag{y}[c]{Throughput (bits/s)}
    \psfrag{x}[c]{Duty-cycle (in percentage)}
	\includegraphics[scale=0.875]{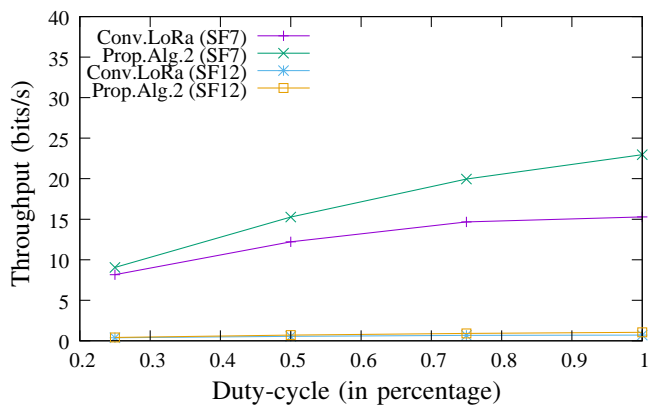}
	\caption{Colliding signals negatively impact the throughput in LoRa. However, our algorithm is able to increase the throughput by up to 25\% even for synchronized signals and 100 EDs.}
	\label{figures:throughputSync}
\end{figure}

Figure~\ref{figures:throughputSync} shows the throughput as a function of the duty-cycle. The throughput computed by Conv.LoRa shows the same performance as the one computed for the desynchronized signals. However, with our second algorithm, the gateway is able to decode one frame for each collision of two frames, provided that the other frame is retransmitted. Thus, our algorithm computes a gain of up to 25\% compared to LoRa. Compared to the case where transmissions are slightly desynchronized, we observe a decrease of 50\% of the throughput.

\section{Conclusion}
\label{section:conclusion}

Collisions in LoRa negatively impact the throughput of the network, which is already very limited by definition. In this paper, we propose two novel collision resolution algorithms that enable to decode some cases of collisions in LoRa by exploiting the specific properties of this physical layer. Our first algorithm focuses on the case where nodes are slightly desynchronized. The second algorithm focuses on the case where nodes are synchronized. Based on our simulation results, we observe that the first algorithm is able to significantly improve the throughput, by decoding all collisions of two signals, and many collisions of three signals. The second algorithm is also able to improve the throughput, but by decoding only one signal when two signals are colliding. These results promote the investigations of a new MAC layer based on LoRa, leveraging on the proposed collision resolution algorithms and thereby outperforming LoRaWAN.

\bibliographystyle{IEEEtran}

\end{document}